\documentclass{article}
\usepackage{url}
\usepackage{graphicx}
\usepackage{amsmath,amsfonts,amssymb}
\usepackage{epsfig,times,latexsym,enumerate,lscape}
\usepackage{amsmath,amssymb,amsfonts}
\usepackage{graphicx}
\usepackage{textcomp}
\usepackage{pgfplots}
\pgfplotsset{/pgfplots/ybar legend/.style={
    /pgfplots/legend image code/.code={%
       \draw[##1,/tikz/.cd,yshift=-0.25em]
        (0cm,0cm) rectangle (3pt,0.8em);},
   },
}
\usepackage{tikz}
\definecolor{trans}{HTML}{b4b4a9}

\begin{document}
\title{\large\bf Privacy-Preserving Mutual Authentication and Key Agreement Scheme for Multi-Server Healthcare System}
\author{Trupil Limbasiya, \footnote{BITS, Pilani, Dept. of CS \& IS, Goa Campus, Goa, India, Email: p20170417@goa.bits-pilani.ac.in} \hspace{0.15mm} Sanjay K. Sahay, \footnote{BITS, Pilani, Dept. of CS \& IS, Goa Campus, Goa, India, India, Email: ssahay@goa.bits-pilani.ac.in} \hspace{0.15mm} and  Bharath Sridharan\footnote{BITS, Pilani, Dept. of CS \& IS, Goa Campus, Goa, India, India, Email: bharathssun@gmail.com}}

\date{}

\maketitle

\begin{abstract}
	The usage of different technologies and smart devices helps people to get medical services remotely for multiple benefits. Thus, critical and sensitive data is exchanged between a user and a doctor. When health data is transmitted over a common channel, it becomes essential to preserve various privacy and security properties in the system. Further, the number of users for remote services is increasing day-by-day exponentially, and thus, it is not adequate to deal with all users using the one server due to the verification overhead, server failure, and scalability issues. Thus, researchers proposed various authentication protocols for multi-server architecture, but most of them are vulnerable to different security attacks and require high computational resources during the implementation. To Tackle privacy and security issues using less computational resources, we propose a privacy-preserving mutual authentication and key agreement protocol for a multi-server healthcare system. We discuss the proposed scheme's security analysis and performance results to understand its security strengths and the computational resource requirement, respectively. Further, we do the comparison of security and performance results with recent relevant authentication protocols.
\vspace*{0.1cm}
~\\
{\it Mutual Authentication, Privacy, Multi-Server, Healthcare Data}
\end{abstract}

\section{Introduction}
One of the emerging fields in the fast-developing world is the Internet of Things (IoT), which is a network system of interrelated devices to sense, compute, and share meaningful information with other heterogeneous devices. The IoT system's impact is highly increased in society due to advanced computing and data availability around the world \cite{Atzori2010}. Thus, it is used in various applications such as intelligent transportation, smart agriculture, home/industry automation, smart grids, smart healthcare, and relevant applications \cite{Zanella2014}. With smart devices and technologies, today's healthcare system is enriched for better healthcare services. Thus, it has various applications like regular patient monitoring, early diagnostics and prediction, health assistant/feedback, medication management, and emergency alerts \cite{Islam2015}. 

\begin{figure}[!h]
\centering
\includegraphics[width=75mm]{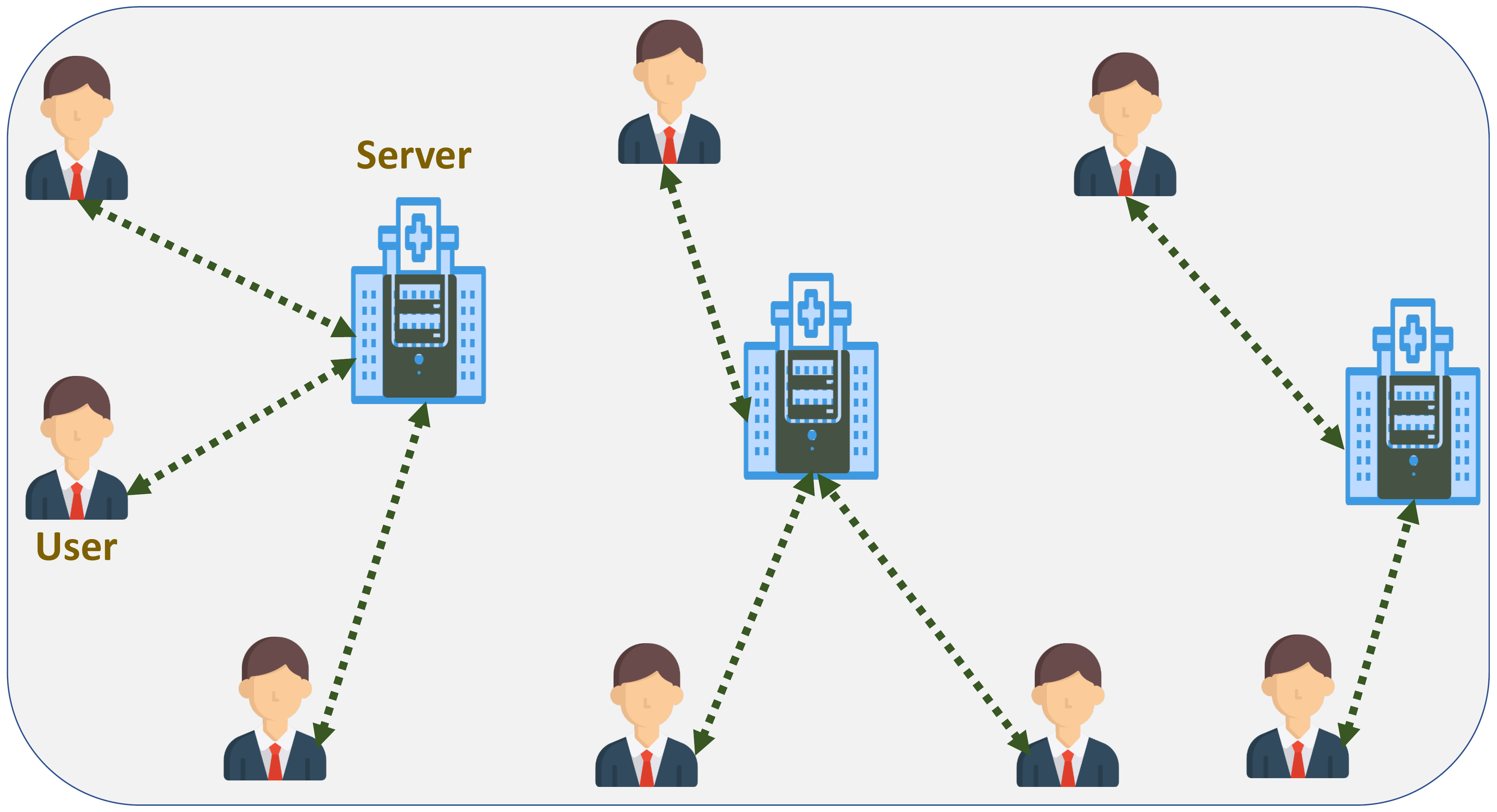}
\caption{The overview of multi-server architecture for healthcare applications }
\label{fig1}
\end{figure}

Health data is sent from a patient to the server for various processes in a patient's interest. Thus, the server is only responsible for providing various services to all registered users in the single-server architecture, and thus, it creates an issue of service availability for authorized users if the server fails due to some reason. Further, it also increases the verification overhead at the server before providing services to the requested users. Besides, the scalability problem is present in single-server architecture. Hence, the multi-server architecture came into the picture for uninterrupted services with more efficient and convenient legal users \cite{Lwamo2019}. Fig. \ref{fig1} shows the multi-server architecture for the healthcare system in which users (i.e., patients) and the hospital server are connected to transmit vital information and get important services quickly. Initially, users and servers should register with the registration center before starting legal services in the future. 

In the multi-server architecture, it is essential to achieve different requirements such as less computation and communication cost, only one-time user registration, session key agreement, and mutual authentication \cite{Juang2004}. Communication technology is daily expanding, and data is sent over a public channel. Thus, there is a need to authenticate the users and their access to IoT applications and systems to resist various security and privacy attributes like data modification, user/server impersonation, message delay, information disclose, user anonymity, and message re-transmission \cite{Yang2017}. To protect the sensitive and confidential health data over a public channel and illegal services from the server, researchers designed various authentication protocols for the Internet of things (IoT) and healthcare systems \cite{Ferrag2017}, \cite{Masdari2017}. However, most of the various authentication mechanisms for the multi-server architecture are insecure against different attacks like modification, denial of service, server/user impersonation, replay, man-in-the-middle, password guessing, stolen smart card, and session key disclosure. The computational resource requirement is also high in the execution cost, communication cost, and storage cost. Therefore, it is required for the healthcare system in which users and servers can connect to exchange vital information for various medical services by preserving user privacy, authentication, confidentiality, and integrity \cite{Li2010}, \cite{Ameen2012JMS}. 

Our Contributions: We propose a privacy-preserving authentication and key agreement mechanism for a multi-server healthcare system to tackle the security, privacy, and performance issues in the multi-server architecture.

\begin{itemize}
	\item Design a secure and privacy-preserving authentication and key agreement scheme using SHA-256 to reduce the requirement of computation resources.
	\item Discuss security proof and analysis of the proposed protocol to know its security strengths against various relevant security attributes.
	\item Analyze the test-bed performance results of the proposed mechanism for the execution time, communication cost, and storage cost. Further, we do the comparison of the proposed scheme with related authentication and key agreement protocols.
\end{itemize}

\textit{Paper structure}: In Section 2, we discuss different features and defects of the existing relevant authentication schemes. Section 3 presents preliminaries by explaining the system model and the adversary model. In Section 4, we propose a privacy-preserving authentication and key agreement scheme for a multi-server healthcare system. Section 5 discusses the formal security proof and security analysis to understand the proposed protocol's security strengths. Section 6 explains the proposed scheme's various performance results and compares them with the recent relevant authentication schemes. We conclude our work in Section 7.

\section{Related Works}
In 2009, Liao et al. \cite{Liao2009} suggested a dynamic identity based remote user authentication scheme using a one-way hash for multi-server environments, offering user password update without involving any trusted party. They claimed that their proposal provides two-factor mutual authentication and user anonymity, resisting replay, server spoofing, forward secrecy, stolen verifier, and insider attacks. According to \cite{Hsiang2009}, the scheme \cite{Liao2009} is insecure to insider, masquerade, server spoofing, and registration center spoofing attacks.

In 2011, Sood et al. \cite{Sood2011} showed that stolen smart card, impersonation, and replay attacks are possible in \cite{Hsiang2009}. To deal with these attacks, they \cite{Sood2011} presented a dynamic identity-based authentication scheme for multi-server architecture using smart cards. However, Sood's implementation itself is unsafe from leak-of-verifier and stolen smart card attacks, as discussed in \cite{Li2012}. Moreover, it does not offer session key agreement and mutual authentication. Li et al. \cite{Li2012} proposed a dynamic identity-based authentication scheme using ring/group signatures to deal with previous security challenges. However, the scheme in \cite{Li2012} is not adequate for multi-server environments because of the control server necessity during the authentication phase, and it cannot withstand to eavesdropping, denial of service, and impersonation attacks as discussed in \cite{Mishra2014Expert}.

In 2012, Tsaur et al. \cite{Tsaur2012} presented an authentication and key agreement mechanism using time-stamp and clock synchronization concepts to resist a replay attack, but it does not achieve user anonymity as the user identity is sent in plain-text over a public channel. Further, the scheme \cite{Tsaur2012} is vulnerable to insider, forward secrecy, and known plain-text attacks according to \cite{Mishra2014Expert}, \cite{Xue2014}. Li et al. \cite{Li2013} proposed an extended authentication and key agreement mechanism to withstand against multi-user login and impersonation attacks by preserving user anonymity. However, it is still insecure to an insider attack and cannot achieve forward secrecy and user anonymity.

In 2014, Xue et al. \cite{Xue2014} proposed an anonymous dynamic pseudonym identity-based authentication and key agreement scheme using non-tamper-resistance assumptions about the smart cards. To support the local user password update, the password change is done locally and without interacting with the remote server, but it creates a problem to verify the old password's authenticity before updating the new password.  Lee et al. \cite{Lee2014} proposed a two-factor authentication and key agreement protocol based on extended chaotic maps in multi-server environments, but the stored data in a smart card could be extracted when it is lost or stolen, and it does not achieve user anonymity. In 2015, Chen et al. \cite{Chen2015} presented a two-factor authentication scheme with anonymity for multi-server environments based on symmetric cryptographic primitives, and it is verified using Burrows-Abadi-Needham (BAN) logic, privileged-insider, and user impersonation attacks.

Odelu et al. \cite{Odelu2015} proposed a biometric-based authentication protocol using a smart card and Elliptic Curve Cryptography (ECC) for multi-server architectures and stated that their scheme could resist passive and active attacks with less communication cost, computational cost. However, it is vulnerable to impersonation and smart card lost attacks. Lu et al. \cite{Lu2015} proposed an authentication and key agreement scheme using biometric identity for multi-server environments, but it does not fulfill forward secrecy and two-factor security. Irshad et al. \cite{Irshad2017} proposed a multi-server authentication and key agreement mechanism using biometric identity for more efficient performance, but it does not provide smart card revocation and mutual authentication when the registration center is offline. Further, it is insecure to user impersonation and password guessing attacks.

In 2018, Ji et al. \cite{Ji2018} presented a blockchain-based multi-lever location sharing scheme to achieve decentralization, multi-level privacy protection, confidentiality, and unforgeability. However, it does not confirm under which critical condition the patient's location data will be retrieved. Mishra et al. \cite{Mishra2018} suggested an authenticated key agreement scheme to preserve user privacy and resist against various security attacks. However, the scheme in \cite{Mishra2018} is insecure to denial of service (DoS), server impersonation, and replay attacks as the time-stamp verification is not done during the control server to the $j^{th}$ server and $j^{th}$ server to a user. Besides, the control server works as an intermediary to verify the user request in the multi-server system.

In 2019, Qiao et al. \cite{Qiao2019} presented a chaotic map-based authenticated key agreement scheme (AKAS) for strong anonymity in a multi-server environment to improve security. However, the main drawback of chaotic encryption algorithms is that they use floating-point calculations. It practically makes software/hardware implementation inefficient and complex compared to the traditional ciphers such as AES and DES (as they only operate with integer operations). Moreover, it is vulnerable to server impersonation, stolen smart card, password guessing, replay, and insider attacks.

Authors \cite{Limbasiya2019} suggested a key agreement mechanism for multi-server architecture to resist against impersonation, insider, man-in-the-middle, password guessing, session key disclosure, smart card lost, and replay attacks. Further, the performance results are comparatively better in \cite{Limbasiya2019}. Presently, user privacy is necessary for the healthcare system because the patient's critical and sensitive information is involved while accessing medical services remotely. Therefore, it also becomes essential to preserve user privacy, authentication, confidentiality, and integrity while designing an authentication protocol for a multi-server healthcare system.

\section{Preliminaries}
We describe the system model and the attacker model to get the general overview of the proposed scheme and the capability of an adversary.

\subsection{System Model}
\begin{figure}[!h]
\centering
\includegraphics[width=70mm]{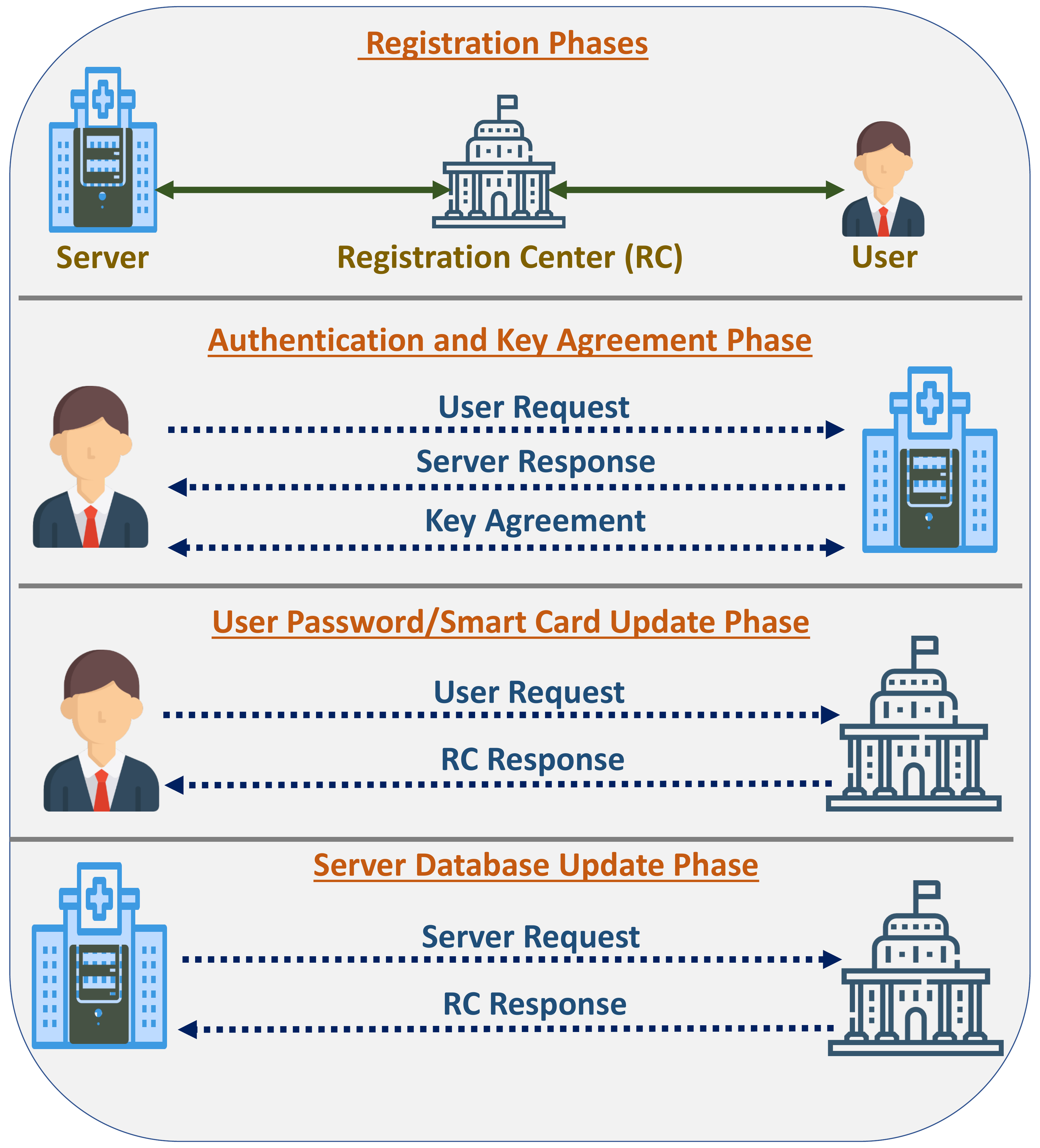}
\caption{The system model for the proposed protocol}
\label{systemmodel}
\end{figure}

We firstly illustrate the system model to understand the basic outline of the proposed mechanism for the multi-server architecture. The proposed system includes three entities (the registration center (RC), servers, and users), and the role of each entity is explained as follows. The flow diagram is shown in Fig. \ref{systemmodel} in which different phases are illustrated for the proposed protocol. The server registration phase is performed between a server and the RC, whereas the user registration phase is executed between a user and RC. Thus, only registered users can get services from authorized servers after the mutual authentication. If the user wants to update his/her password or smart card, it can be done through the user password/smart card update phase by confirming a user's legality. All authorized servers regularly connect with the RC to get the up-to-date database of all enrolled users so they can verify the received requests and provide services to the legitimate users.

\begin{enumerate}
	\item \textbf{The registration center (RC)} is the trusted authority to regularly register users and servers and update its secure database based on the registration. Besides, the RC connects with the existing registered users and servers to provide the list of newly enrolled servers and users so that legal users can get services from the servers' extensive network.
	
	\item \textbf{Server} initially registers with the RC to become an authorized service provider in the system and get the list of users to know the registered users with the RC so a server can provide services to users around the world. A registered server (e.g., hospital) is considered as the trusted party.
	
	\item \textbf{User} registers with the RC to be a legitimate user to get services from different servers across the world. During the registration phase, a user (e.g., patient) gets the list of legal servers so s/he can connect for various services from different authorized servers.
	
\end{enumerate}

\subsection{Adversary Model}
We consider the following valid security and performance assumptions by referring \cite{Kocher1999}, \cite{Messerges2002}, \cite{Madhusudhan2012}, \cite{Limbasiya2017} to understand the capability of an adversary.

\begin{enumerate}

	\item Servers and users are registered using a secure infrastructure as the transport layer security (TLS), and data is sent over a public channel during the authentication and key agreement phase. Therefore, an adversary ($\mathcal{A}$) can capture all transmitted messages via a public channel, but s/he cannot access any sent messages over a secure channel.
	
	\item $\mathcal{A}$ can stop, reroute, delay, resend, modify, or delete the transferred messages over a public channel.
		
	\item If $\mathcal{A}$ gets a smart card of a registered user, then s/he can extract saved parameters from a smart card through power analysis.
	
	\item A registered user ($U_{i}$) may act as an adversary for another legal user ($U_{j}$). If $U_{i}$ can establish a connection with the server behalf of $U_{j}$, then $U_{i}$ is an adversary for $U_{j}$. Hence, $U_{i}$ can play two roles (as a legitimate user and an adversary) in the system. 
	
	\item $\mathcal{A}$ can guess only one value at a time to know another unknown parameter in the system. It describes that $\mathcal{A}$ cannot guess more than one parameter in polynomial time.
		
	\item We consider $\mathcal{I} = \mathcal{J} \oplus \mathcal{K}$. If $\mathcal{A}$ knows $\mathcal{J}$ and $\mathcal{K}$, then only s/he can get $\mathcal{I}$, but $\mathcal{A}$ cannot compute $\mathcal{I}$, $\mathcal{J}$ or $\mathcal{K}$ by knowing only one parameter ($\mathcal{J}$ or $\mathcal{K}$ or $\mathcal{I}$).
			
\end{enumerate}

\section{The Proposed Protocol}
We propose a privacy-preserving authentication and key agreement protocol for a multi-server healthcare system. In this scheme, a user once registers with the RC to become a legal user to get services from different authorized servers legitimately. Besides, a server should also perform the registration procedure one time with the RC, and after that, that server is permitted to provide services to legitimate users. The authentication and key agreement phase is performed between a user and a server to verify each other mutually to exchange vital medical information over a public channel. The proposed protocol consists of five phases as (i) server registration (ii) user registration (iii) authentication and key agreement (iv) user password/smart card update (v) server database update. We explain each phase in-detail as follows. We use different notations in the proposed scheme, and they are described in Table \ref{notation}. Since the proposed scheme is based on the multi-server architecture, the system includes multiple users and servers, where a user is denoted as $U_{i}$ (where $i= 1, 2, 3, ...$), and a server is as $S_{j}$ (where $i= 1, 2, 3, ...$).  Here, there is no correlation between the index $i$ and $j$. It means that any user can connect with any server.

\begin{table}[!h]
\caption{Notations Used in the Scheme}
\label{notation}
\begin{tabular}{|l|l|}
\hline
\textbf{Notation} & \textbf{Explanation}\\ \hline
	$U_{i}$  & A user \\
	$S_{j}$  & A Server \\
	$ID_{i}$/$PW_{i}$ & Identity/Password of $U_{i}$ \\
	$SC_{i}$ & Smart card of $U_{i}$ \\
	$UID_{i}$ & Anonymous identity of $U_{i}$ \\
	$List_{UID_{i}}$ & A list of $UID_{i}$ \\
	$List_{C_{i}}$ & A list of $C_{i}$ \\
	$List_{S_{j}}$ & A list of $S_{j}$ \\
	$ID_{j}$/$PW_{j}$ & Identity/Password of $S_{j}$\\
	$SSK_{j}$ & Server service key of $S_{j}$\\
	$SRT_{j}$ & Server registration time-stamp for $S_{j}$\\
	$USK_{i}$ & User service key of $U_{i}$\\
	$Loc_{j}$ & Location of $S_{j}$\\
	$K_{RC}$ & The registration center 256-bit key\\
	$r_{S}$/$r_{1}$/$r_{2}$/$r_{3}$ & Random nonce \\
	$T_{1}$/$T_{3}$ & Current time-stamp at $U_{i}$ side \\
	$T_{2}$ & Current time-stamp at $S_{j}$ side \\
	$\Delta T$ & Threshold time\\
	$SK_{ij}$ & The session key between $U_{i}$ and $S_{j}$ \\
	$VT_{ij}$ & Validity of $SK_{ij}$ between $U_{i}$ and $S_{j}$ \\
\hline
\end{tabular}
\end{table}

\subsection{Server Registration Phase}
\begin{figure}[!h]
\centering
\includegraphics[width=81mm]{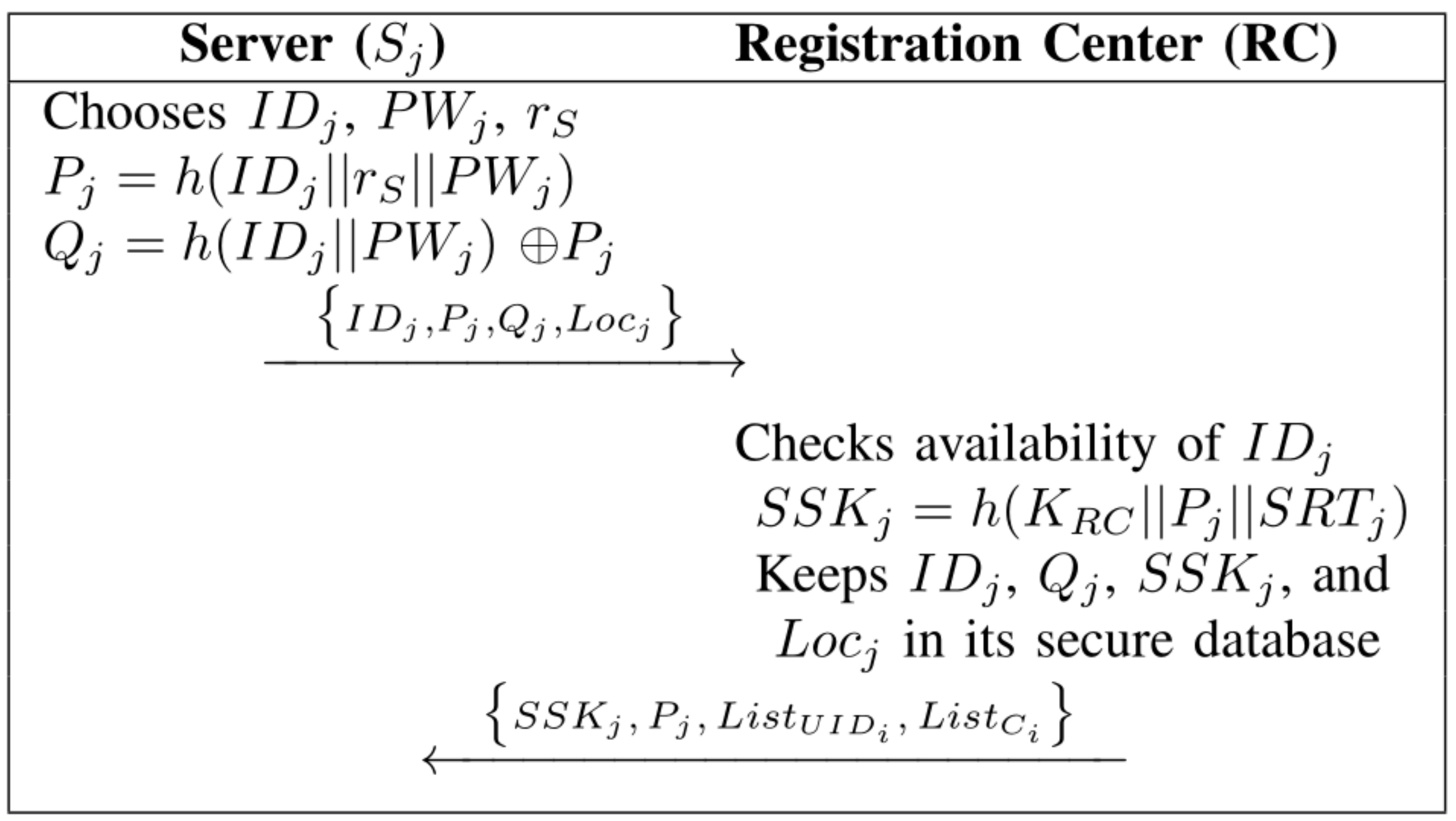}
\caption{The proposed server registration phase}
\label{SEREG}
\end{figure}

Only authorized service providers are allowed to provide various services legally to users, and thus, a server performs the following steps. It is also shown in Fig. \ref{SEREG}.

\begin{enumerate}
	\item A server ($S_{j}$) chooses its $ID_{j}$, $PW_{j}$, $r_{S}$ to calculate $P_{j}=h(ID_{j}||r_{S}||PW_{j})$, $Q_{j}= h(ID_{j}||PW_{j}) \oplus P_{j}$ and sends \big\{\text{$ID_{j}$, $P_{j}$, $Q_{j}$, $Loc_{j}$}\big\} to the RC through the TLS protocol \cite{Kocher1999}. Here, $P_{j}$ and $Q_{j}$ are computed parameters for $S_{j}$.
	
	\item The RC confirms the availability of $ID_{j}$ in its secure database. If $ID_{j}$ is not present in the database, then the RC computes $SSK_{j}=h(K_{RC}||P_{j}||SRT_{j})$ and keeps $ID_{j}$, $SSK_{j}$, $Q_{j}$, and $Loc_{j}$ in its secure database. Further, the RC saves $SSK_{j}$, $P_{j}$, $List_{UID_{i}}$, and $List_{C_{i}}$ in a tamper-resistant memory and provides it to $S_{j}$ through the TLS protocol to complete the server registration legitimately.
	
\end{enumerate}

\subsection{User Registration Phase} \label{UserRegistrationPhase}
\begin{figure}[!h]
\centering
\includegraphics[width=81mm]{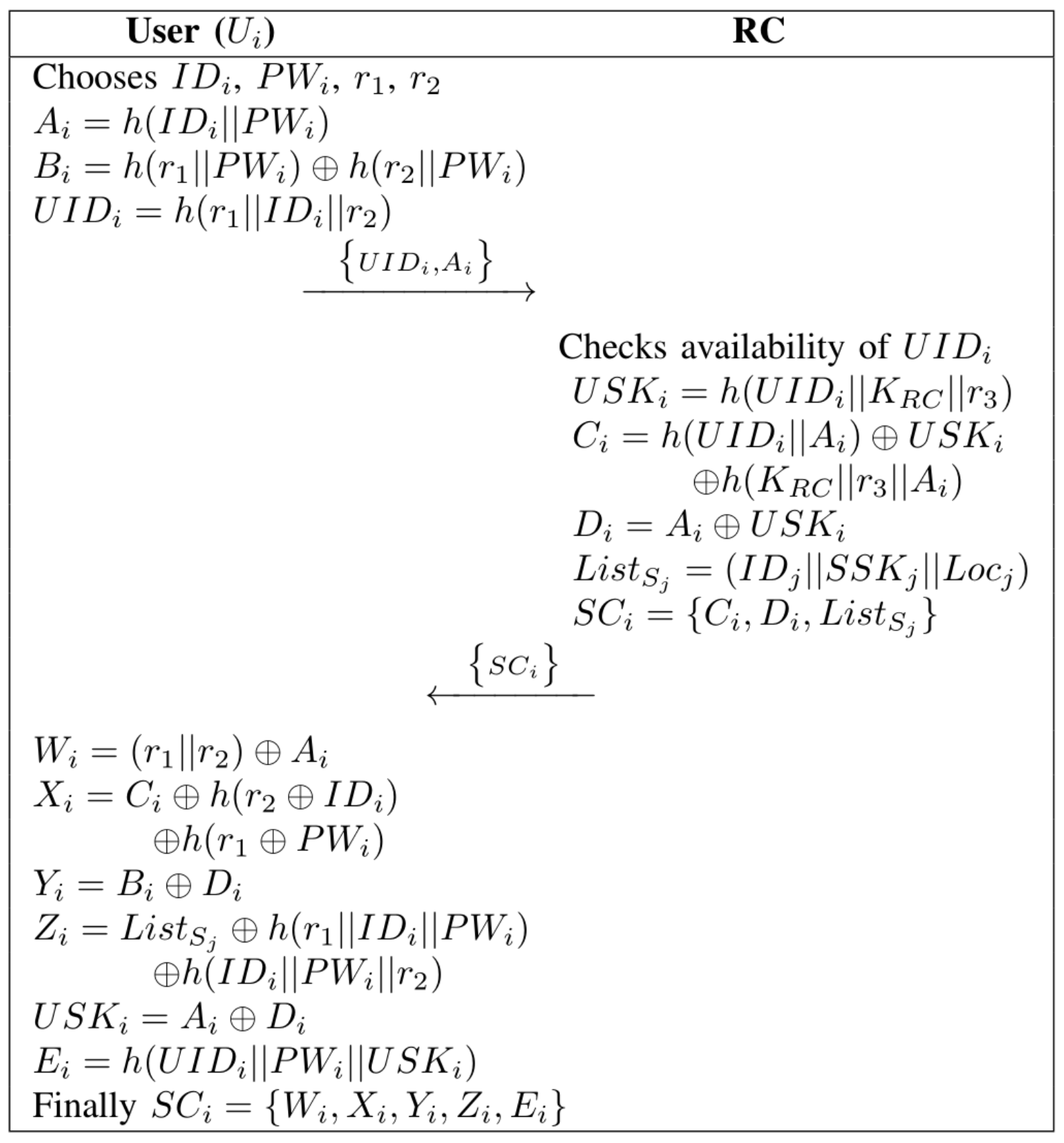}
\caption{The proposed user registration phase}
\label{USREG}
\end{figure}

A user ($U_{i}$) should register with the RC once to get various services from different servers and complete the registration process, $U_{i}$ executes the following steps. This phase is also displayed in Fig. \ref{USREG}.

\begin{enumerate}
	\item $U_{i}$ selects his/her $ID_{i}$, $PW_{i}$, $r_{1}$, $r_{2}$ to calculate $A_{i}=h(ID_{i}||PW_{i})$, $B_{i}=h(r_{1}||PW_{i}) \oplus h(r_{2}||PW_{i})$, $UID_{i}$ $=h(r_{1}||ID_{i}||r_{2})$ and sends \big\{$UID_{i}, A_{i}$\big\} to the RC over a secure infrastructure \cite{Kocher1999}.
	
	\item The RC checks the availability of $UID_{i}$. If it is, then then RC calculates $USK_{i}=h(UID_{i}||K_{RC}||r_{3})$, $C_{i}=h(K_{RC}||r_{3}||A_{i}) \oplus USK_{i} \oplus h(UID_{i}||A_{i})$, $D_{i}=A_{i} \oplus USK_{i}$, $List_{S_{j}} = (ID_{j}||SSK_{j}||Loc_{j})$, where $C_{i}$ and $D_{i}$ are calculated parameters for $U_{i}$, and $List_{S_{j}}$ contains various values of different registered servers. Further, the RC saves $UID_{i}$, $C_{i}$ in its secure database and $C_{i}$, $D_{i}$, $List_{S_{j}}$ in the smart card ($SC_{i}$) of $U_{i}$. After that, the RC provides $SC_{i}$ to $U_{i}$.
	
	\item $U_{i}$ computes $W_{i}= (r_{1}||r_{2}) \oplus A_{i}$, $X_{i} = h(r_{2} \oplus ID_{i}) \oplus C_{i} \oplus h(r_{1} \oplus PW_{i})$, $Y_{i} = B_{i} \oplus D_{i}$, $Z_{i} = h(r_{1}||ID_{i}||PW_{i})$ $\oplus List_{S_{j}} \oplus h(ID_{i}||PW_{i}||r_{2})$, $USK_{i} = A_{i} \oplus D_{i}$, $E_{i} = h(UID_{i}||PW_{i}||USK_{i})$. Finally, $U_{i}$ removes $C_{i}$, $D_{i}$, $List_{S_{j}}$ from $SC_{i}$ and stores $W_{i}$, $X_{i}$, $Y_{i}$, $Z_{i}$, $E_{i}$ in $SC_{i}$ to complete the registration process.
	
\end{enumerate}

\begin{figure*}[!h]
\centering
\includegraphics[width=140mm]{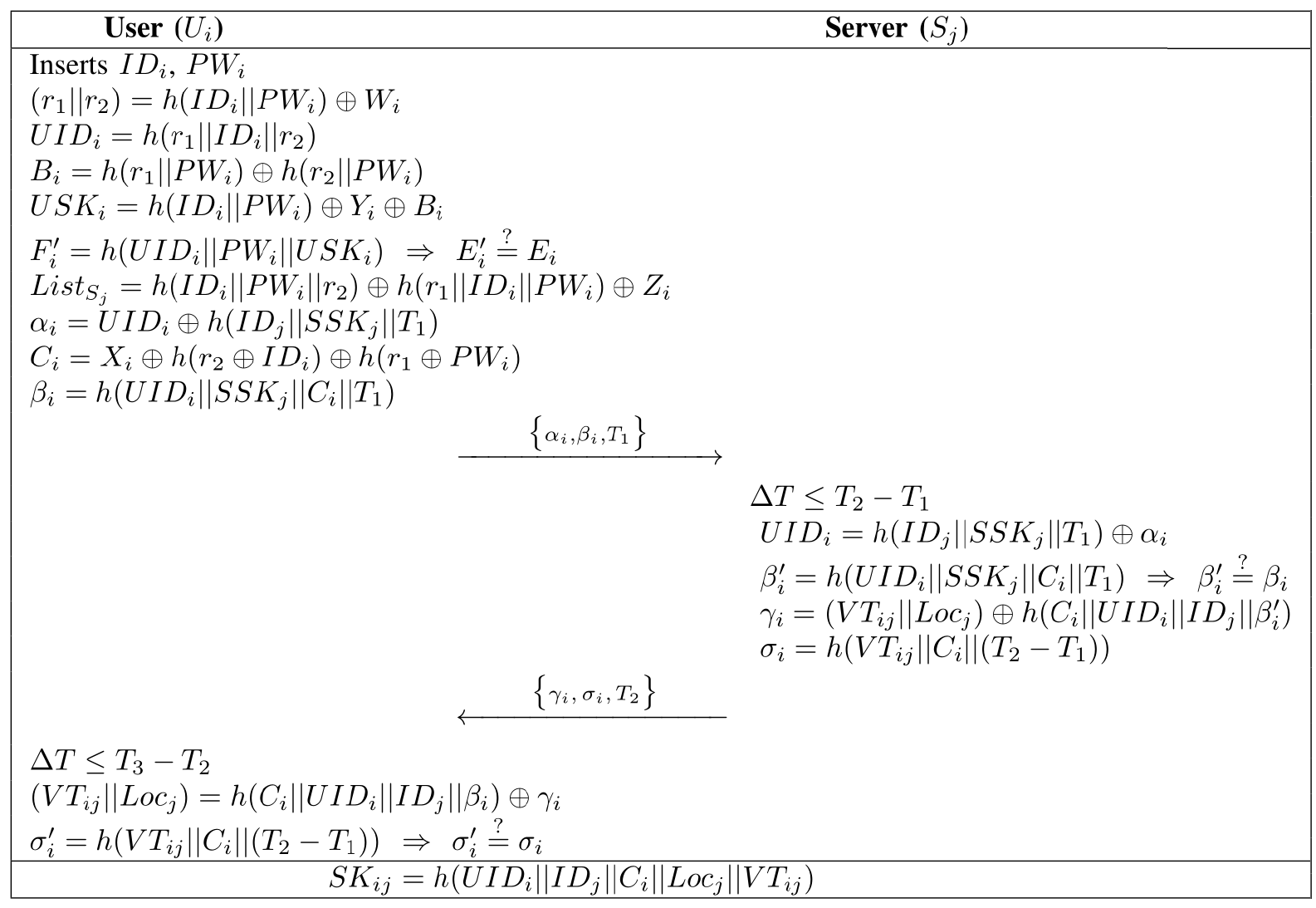}
\caption{The proposed authentication and key agreement phase}
\label{AUKA}
\end{figure*}

\subsection{Authentication and Key Agreement Phase} \label{AuthenticationKAPhase}
When a user ($U_{i}$) is interested to get legal services from an authorized server ($S_{j}$), s/he computes a login request to send it to that server via a public channel to verify his/her authenticity. After that, $S_{j}$ confirms the freshness and legality of the received request. If it holds, then $S_{j}$ calculates a response message for the received request to send it to $U_{i}$ over a common channel. Now, $U_{i}$ confirms the authenticity of $S_{j}$ for mutual authentication. If it is valid, then both ($U_{i}$ and $S_{j}$) computes the common session key to start a connection, and this key is valid for a temporary period. If the session key gets expired, then both should recompute the session key using new credentials. The authentication and key agreement phase is also represented in Fig. \ref{AUKA}.

\begin{enumerate}
	
	\item $U_{i}$ inserts $ID_{i}$, $PW_{i}$ to calculate $(r_{1}||r_{2})=W_{i} \oplus h(ID_{i}||PW_{i})$, $B_{i}=h(r_{1}||PW_{i}) \oplus h(r_{2}||PW_{i})$, $USK_{i}$ $=h(ID_{i}||PW_{i}) \oplus Y_{i} \oplus B_{i}$,	$UID_{i}=h(r_{1}||ID_{i}||r_{2})$, $F'_{i}=h(UID_{i}||PW_{i}||USK_{i})$ and confirms the user's legitimacy by $E'_{i}\stackrel{?}{=}E_{i}$.
	
	\item If $E'_{i}=E_{i}$, $U_{i}$ calculates $List_{S_{j}} =h(ID_{i}||PW_{i}||r_{2})$ $\oplus Z_{i} \oplus h(r_{1}||ID_{i}||PW_{i})$, $\alpha_{i}=h(ID_{j}||SSK_{j}||T_{1}) \oplus UID_{i}$, $C_{i}=X_{i} \oplus h(r_{2} \oplus ID_{i}) \oplus h(r_{1} \oplus PW_{i})$, and $\beta_{i}=h(UID_{i}||SSK_{j}||C_{i}||T_{1})$ to send \big\{$\alpha_{i}, \beta_{i}, T_{1}$\big\} to $S_{j}$ via a public channel.
	
	\item $S_{j}$ checks the freshness of \big\{$\alpha_{i}, \beta_{i}, T_{1}$\big\} by comparing $\Delta T \leq T_{2} - T_{1}$. If it is valid, then $S_{j}$ computes $UID_{i}=h(ID_{j}||SSK_{j}||T_{1}) \oplus \alpha_{i}$ and $\beta'_{i}=h(UID_{i}||SSK_{j}||C_{i}||T_{1})$ to verify the legality of $U_{i}$ by comparing $\beta'_{i}\stackrel{?}{=}\beta_{i}$. If it holds, then $S_{j}$ calculates $\gamma_{i} = (VT_{ij}||Loc_{j}) \oplus h(C_{i}||UID_{i}||ID_{j}||\beta'_{i})$, $\sigma_{i} = h(VT_{ij}||C_{i}||(T_{2}-T_{1}))$ to send \big\{\text{$\gamma_{i}, \sigma_{i}, T_{2}$}\big\} to $U_{i}$ via a common channel.
	
	\item $U_{i}$ confirms the freshness of \big\{\text{$\gamma_{i}, \sigma_{i}, T_{2}$}\big\} through $\Delta T$ $\leq T_{3} - T_{2}$. If valid, $U_{i}$ calculates $(VT_{ij}||Loc_{j})= \gamma_{i} \oplus h(C_{i}||UID_{i}||ID_{j}||\beta_{i})$ and $\sigma'_{i} = h(VT_{ij}||C_{i}||(T_{2}-T_{1}))$ to check $\sigma'_{i}\stackrel{?}{=}\sigma_{i}$. If $\sigma'_{i} \neq \sigma_{i}$, then $U_{i}$ ends the session immediately.
	
	\item If $\sigma'_{i}=\sigma_{i}$, then only $U_{i}$ and $S_{j}$ compute $SK_{ij}=h(UID_{i}||ID_{j}||C_{i}||Loc_{j}||VT_{ij})$ as the common session key using mutual credentials to start a connection for medical services, and this session key is temporary and valid for a fixed time period ($VT_{ij}$).
	
\end{enumerate}

\subsection{User Password/Smart Card Update Phase}	\label{PasswordSmartCardUpdatePhase}
When $U_{i}$ wants to update his/her password or smart card (to get the list of newly registered servers), s/he should proceed to the following steps.

\begin{enumerate}
	\item This step is the same as Step-1 of Section \ref{AuthenticationKAPhase}.
	
	\item If $E'_{i}=E_{i}$, then $U_{i}$ computes $C_{i}= h(r_{2} \oplus ID_{i}) \oplus X_{i} \oplus h(r_{1} \oplus PW_{i})$, $\tau_{i}=h(C_{i}||T_{4}||UID_{i})$ to send \big\{\text{$UID_{i}, \tau_{i}, T_{4}$}\big\} to the RC.
	
	\item The RC checks the freshness of \big\{\text{$UID_{i}, \tau_{i}, T_{4}$}\big\} by $\Delta T \leq T_{5} - T_{4}$. If valid, then the RC calculates $\tau'_{i}=h(C_{i}||T_{4}||UID_{i})$ to verify $\tau'_{i}\stackrel{?}{=}\tau_{i}$. If it holds, then only the RC provides $List_{S_{j}}$ to $U_{i}$.
	
	\item $U_{i}$ computes $Z^{New}_{i}=List_{S_{j}} \oplus h(r_{1}||ID_{i}||PW_{i}) \oplus h(ID_{i}||PW_{i}||r_{2})$ and replaces $Z_{i}$ by $Z^{New}_{i}$ in $SC_{i}$.

\end{enumerate}

\subsection{Server Database Update Phase}	\label{ServerDatabaseUpdatePhase}
A server ($S_{j}$) regularly updates its database to have the latest list of registered users and for this, $S_{j}$ performs as follows through a secure infrastructure.

\begin{enumerate}
	\item $S_{j}$ inserts $ID_{j}$, $PW_{j}$ to calculate $Q_{j}= h(ID_{j}||PW_{j})$ $\oplus P_{j}$, $\omega_{j}=h(Q_{j}||T_{6}||SSK_{j})$ and sends a database update request as \big\{\text{$ID_{j}, \omega_{j}, T_{6}$}\big\} to the RC.
	
	\item The RC checks its freshness through $\Delta T \leq T_{7} - T_{6}$. If it holds, the RC computes $\omega'_{j}=h(Q_{j}||T_{6}||SSK_{j})$ to confirm $\omega'_{j}\stackrel{?}{=}\omega_{j}$. If both are equal, then the RC securely updates the server database with the newly registered users and their credentials.
	
\end{enumerate}

\section{Security Assessment of the Proposed Scheme}
We describe the security proof and analysis for the proposed scheme to confirm its security and privacy robustness against various pertinent attacks and adversaries.

\subsection{Security Proof}
We show the formal security proof based on the random oracle model (ROM) for the proposed scheme to understand its security strengths against the capabilities of insider and external adversaries, as follows. A game is played between the challenger ($\mathcal{C}$) and an adversary ($\mathcal{A}$) to decide the non-negligible possibility to win a game polynomially by $\mathcal{A}$ for the given challenge by $\mathcal{C}$. The game is outlined for the proposed authentication and key agreement (refer Section \ref{AuthenticationKAPhase}), user password/smart card update (refer Section \ref{PasswordSmartCardUpdatePhase}), and server database update (refer Section \ref{ServerDatabaseUpdatePhase}) phases.

\textit{Authentication and Key Agreement \textbf{$\mbox{---}$} Oracle\textbf{:}} $\mathcal{A}$ sends \{$\alpha_{i}$, $\beta_{i}$, $T_{1}$\} to $\mathcal{C}$ to approve the sent bogus values. Here, $\mathcal{C}$ is a legal server ($S_j$). $\mathcal{C}$ checks the validity of the received parameters through $\Delta T$ and legality by $\beta'_{i}\stackrel{?}{=}\beta_{i}$. If both hold at $S_j$ side, then only $\mathcal{C}$ proceeds further to send \{$\gamma_{i}$, $\sigma_{i}$, $T_2$\} to $\mathcal{A}$. Otherwise, $\mathcal{C}$ rejects the session straight away. If $\mathcal{A}$ receives response parameters from $\mathcal{C}$ and can compute the common session key ($SK_{ij}$), then only $\mathcal{A}$ wins the game.

\textit{User Password/Smart Card Update \textbf{$\mbox{---}$} Oracle\textbf{:}} $\mathcal{A}$ transfers \{$UID_i$, $\tau_{i}$, $T_4$\} to $\mathcal{C}$ to update the user password or smart card illegally. Here, $\mathcal{C}$ is the registration center (RC). $\mathcal{C}$ checks the freshness of the obtained values based on $\Delta T$ and legitimacy by confirming $\tau'_{i}$ with $\tau_{i}$. If both conditions are valid at the RC side, then only $\mathcal{C}$ computes $Z^{New}_{i}$ to replace $Z_{i}$ by $Z^{New}_{i}$ in $SC_{i}$. Otherwise, $\mathcal{C}$ directly ends the session without proceeding with the next process. If the RC computes $Z^{New}_{i}$, then only $\mathcal{A}$ can succeed in the game.

\textit{Server Database Update \textbf{$\mbox{---}$} Oracle\textbf{:}} $\mathcal{A}$ sends \{$ID_{j}$, $\omega_{j}$, $T_6$\} to $\mathcal{C}$ to get the updated server database values illicitly, where $\mathcal{C}$ is the registration center (RC). $\mathcal{C}$ checks the freshness of the received request through $\Delta T$ and legitimacy by verifying $\omega'_{j}$ with $\omega_{j}$. If both conditions are satisfied at the RC side, then only $\mathcal{C}$ updates the server database. In other cases, $\mathcal{C}$ terminates the session directly. Here, if the server database is updated then only $\mathcal{A}$ succeeds in the game.

The proposed schemes are designed based on SHA-256, which security is well established by NIST, and it is secure against the polynomial time algorithm \cite{Dang2015}. Hence, on this basis, the security proof by contradiction is shown as follows for the proposed schemes by considering an external adversary and an internal adversary.

\emph{\textbf{Definition 1:}} We consider that an adversary is an external adversary ($\mathcal{AA}$). Thus, s/he can capture common channel parameters (transferred in the authentication, user password/smart card update, and server database update phases). Furthermore, $\mathcal{AA}$ does not have any smart card parameters due to an external adversary. $\mathcal{AA}$ aims (i) to impersonate a legitimate user, (ii) update smart card or user password, and (iii) to get updated server database illegally.

\emph{\textbf{Theorem 1:} The proposed protocol resists to $\mathcal{AA}$'s adaptive illegitimate activities under the one-way hash function consideration polynomially in the ROM.}

\textbf{\textit{Proof}:} $\mathcal{AA}$ wants to connect with an authorized server ($S_j$) illegally to get services from $S_j$. Thus, $\mathcal{AA}$ should compute request values ($\alpha_{i}$ and $\beta_{i}$) as per the proposed protocol.  Since $\alpha_{i}=UID_{i} \oplus h(ID_{j}||SSK_{j}||T_{1})$ and $\beta_{i}=h(UID_{i}||SSK_{j}||C_{i}||T_{1})$, $\mathcal{AA}$ requires $UID_{i}$, $ID_{j}$, $SSK_{j}$, and $C_{i}$ to calculate $\alpha_{i}$ and $\beta_{i}$ again. However, $\mathcal{AA}$ does not have all these essential parameters as an external adversary, leading to a failure for the forged request computation. If $\mathcal{AA}$ sends a request by using earlier request parameters (sent by the original user, $U_{i}$ previously), then $\mathcal{AA}$ fails to clear the freshness test because $\mathcal{C}$ immediately checks the freshness of the obtained request, and previously used time-stamp (i.e., $T_{1}$) is not valid beyond the time duration. If $\mathcal{AA}$ uses $T'_1$ instead of $T_{1}$ to pass the freshness test, then also s/he fails to clear $\beta'_{i}\stackrel{?}{=}\beta_{i}$ because $T_{1}$ is an input parameter in the computation of $\beta_{i}$. Consequently, $\mathcal{AA}$ fails to get any successive response from $\mathcal{C}$ due to erroneous computation based on incorrect parameters. Similarly, $\mathcal{AA}$ cannot calculate $\gamma_{i}$ and $\sigma_{i}$ to forge $U_{i}$ due to unavailability of essential values. Hence, $\mathcal{AA}$ cannot win the game to impersonate $U_{i}$ or $S_{j}$ in the proposed authentication and key agreement scheme.

To forge a request in the proposed user password or smart card update phase, $\mathcal{AA}$ should compute $\tau_{i}~[=h(C_{i}||T_{4}||UID_{i})]$ again, but s/he does not know $C_{i}~[=h(r_{2} \oplus ID_{i}) \oplus X_{i} \oplus h(r_{1} \oplus PW_{i})]$ to calculate $\tau_{i}$. Thus, if $\mathcal{AA}$ sends a new request (computed with forged values), then $\mathcal{C}$ does not send a valid response to $\mathcal{AA}$. In addition, $\mathcal{AA}$ cannot use $T'_{4}$ instead of $T_{4}$ because $\mathcal{C}$ confirms the legitimacy of $\tau_{i}$ with $\tau'_{i}$, and $T_{4}$ is used as one input value in $\tau_{i}$. Consequently, $\mathcal{C}$ rejects the received request directly due to $\tau_{i} \neq \tau'_{i}$ at the RC side. $\mathcal{AA}$ cannot proceed further due to incorrect parameters. Therefore, $\mathcal{AA}$ fails to win the game in the proposed user password or smart card update phase.

To access the updated server database, $\mathcal{AA}$ should compute $\omega_{j}$~$[=h(Q_{j}||T_{6}||SSK_{j})]$ correctly. Thus, s/he requires $Q_{j}$ and $SSK_{j}$, but it is not feasible for $\mathcal{AA}$ to get these essential values anyhow due to unavailability of $ID_{j}$, $PW_{j}$, and $P_{j}$. Hence, if $\mathcal{AA}$ sends a request with forged values, then $\mathcal{C}$ ends the session due to the verification failure (in $\Delta T \leq T_{7} - T_{6}$ and $\omega_{j}\stackrel{?}{=}\omega'_{j}$) based on the obtained request. Accordingly, $\mathcal{C}$ does not proceed for further computation. Thus, $\mathcal{AA}$ cannot win the game in the proposed server database update procedure.

\emph{\textbf{Definition 2:}} An adversary is a registered user ($U_{i+1}$) with the RC, considering an internal adversary ($\mathcal{AB}$). Therefore, $\mathcal{AB}$ has his/her credentials with the smart card parameters. Moreover, s/he can intercept an insecure channel to capture public channel values (exchanged during the authentication, user password/smart card update, and server database update phases). $\mathcal{AB}$ aims for (i) to impersonate a legal user ($U_{i}$), (ii) to change user password or update smart card, and (iii) to know updated server database illegitimately.

\emph{\textbf{Theorem 2:} The proposed scheme can withstand $\mathcal{AB}$'s adaptive malicious activities polynomially while considering a one-way hash function in the ROM.}

\textbf{\textit{Proof}:} To get services illegally from $S_j$, $\mathcal{AB}$ should calculate $\alpha_{i}$ and $\beta_{i}$ according to the proposed protocol (see Section \ref{AuthenticationKAPhase}). Thus, $\mathcal{AB}$ needs $UID_{i}$, $ID_{j}$, $SSK_{j}$, and $C_{i}$. We consider that $\mathcal{AB}$ manages $ID_{j}$ and $SSK_{j}$ as an internal adversary (by referring to Definition  2). However, it is not feasible for $\mathcal{AB}$ to compute/get $UID_{i}$ and $C_{i}$ in polynomial time even though s/he is a registered user because s/he does not know $r_{1}$, $r_{2}$, $PW_{i}$, and $X_{i}$. If $\mathcal{AB}$ sends a request with dummy parameters, then it is rejected by $\mathcal{C}$ because $\beta'_{i}\stackrel{?}{=}\beta'_{i}$ test is not cleared. Thus, $\mathcal{AB}$ does not get any correct response from $\mathcal{C}$. Hence, if $\mathcal{AB}$ attempts to forge $\mathcal{C}$ by using incorrect values, then $\mathcal{C}$ rejects the request immediately. Similarly, $\mathcal{AB}$ fails to compute $\gamma_{i}$ and $\sigma_{i}$ to do illegal activities to impersonate $U_{i}$. Therefore, $\mathcal{AB}$ cannot win the game in the proposed authentication and key agreement phase.

$\mathcal{AB}$ requires $UID_{i}$ and $C_{i}$ to send an illegal request to update the user password or smart card. Since $C_{i}$ is not the same value for different users (as described in Section \ref{UserRegistrationPhase}), $\mathcal{AB}$ cannot use his/her $C_{\mathcal{AB}}$ to impersonate $U_{i}$. As discussed in the previous paragraph, $\mathcal{AB}$ does not have $C_{i}$ and $UID_{i}$ for $U_{i}$. In addition, $\mathcal{AB}$ cannot forge the request due to the verification of $\tau_{i}$ and $T_{4}$ at the RC side as the challenger. Therefore, $\mathcal{AB}$ does not get valid response from $\mathcal{C}$ to succeed in the game.

$\mathcal{AB}$ needs $ID_{j}$ and $\omega_{j}$ to get illegal access of the updated server database. $\mathcal{AB}$ knows $ID_{j}$ as an internal adversary (according to Definition  2), but s/he does not get/calculate $Q_{j}~[=P_{j}\oplus h(ID_{j}||PW_{j})]$ because s/he does not know $P_{j}$ and $PW_{j}$. Since $\mathcal{AB}$ does not have $\omega_{j}$ precisely, s/he does not get valid response from $\mathcal{C}$, leading to a failure for winning a game. Moreover, the RC confirms the freshness (through $\Delta T$) and legality (based on $\omega'_{j}\stackrel{?}{=}\omega_{j}$) of the received request, restricting $\mathcal{AB}$ to send dummy parameters to get correct response. Consequently, $\mathcal{AB}$ cannot do forgery in the proposed server database update phase.

\subsection{Security Attacks Analysis}
We explain that how the proposed system can withstand against various security attacks, as follows.

\begin{enumerate}
	\item \textbf{\textit{User Impersonation}:} If an adversary ($\mathcal{A}$) can send a valid login request behalf of a legitimate user ($U_{i}$) and the server ($S_{j}$) considers it as a legal request by $U_{i}$, then a user impersonation attack is feasible in the system. To apply a user impersonation attack in the proposed scheme, $\mathcal{A}$ should know $UID_{i}$, $ID_{j}$, $SSK_{j}$, $C_{i}$ to compute $\alpha_{i}$, $\beta_{i}$ so s/he can send a request to $S_{j}$. There are two cases for $\mathcal{A}$ as (i) registered user and (ii) non-registered user.
\begin{itemize}
	\item If $\mathcal{A}$ is registered user, then s/he knows $ID_{j}$ and $SSK_{j}$ (from own smart card), but s/he does not know $UID_{i} (=h(r_{1}||ID_{i}||r_{2}))$ and $C_{i} (=X_{i}\oplus h(r_{2}\oplus ID_{i}) \oplus h(r_{1} \oplus PW_{i}))$, where $\mathcal{A}$ does not have any knowledge of $ID_{i}$, $PW_{i}$, $r_{1}$, $r_{2}$ even though as another registered user, and it is not possible to guess all these values.
	
	\item	A non-registered user cannot know $UID_{i}$, $ID_{j}$, $SSK_{j}$, $C_{i}$ anyhow as $\alpha_{i}$, $\beta_{i}$ are computed using one-way hash irreversible function, and $\mathcal{A}$ cannot get $UID_{i}$, $ID_{j}$, $SSK_{j}$, $C_{i}$ from a public channel parameters ($\alpha_{i}$, $\beta_{i}$, $\gamma_{i}$, $\sigma_{i}$) anyhow. 
	
\end{itemize}

In the above discussed both cases, $\mathcal{A}$ cannot compute the required parameters to send a valid login request to $S_{j}$. Therefore, the proposed scheme can withstand a user impersonation attack.

\item \textbf{\textit{Server Impersonation}:} If $\mathcal{A}$ can response correctly to the user login request and a user ($U_{i}$) believes on the received response by an adversary, then the system is weak against a server impersonation attack. If $\mathcal{A}$ wants to perform a server impersonation attack in the proposed scheme, then s/he should compute $\gamma^{\mathcal{A}}_{i}$ and $\sigma^{\mathcal{A}}_{i}$ correctly so $U_{i}$ can believe on the sent parameters ($\gamma^{\mathcal{A}}_{i}$ and $\sigma^{\mathcal{A}}_{i}$) by an adversary. To compute these values, $\mathcal{A}$ requires $ID_{j}$, $SSK_{j}$, $Loc_{j}$, $UID_{i}$, $C_{i}$, $\beta_{i}$. We consider that $\mathcal{A}$ (as a registered user) has $List_{S_{j}}$ from which s/he can get $ID_{j}$, $SSK_{j}$, $Loc_{j}$, but it takes more time to find out these values correctly as there are multiple servers are deployed for services. During this long time, $U_{i}$ does not accept any response messages for $\big\{\alpha_{i}, \beta_{i}, T_{1}\big\}$ due to its freshness concern. Besides, $\mathcal{A}$ does not know $C_{i}$ of $U_{i}$, and the validity of $\beta_{i}$ is expired. For all these reasons, an adversary cannot perform a server impersonation attack on the sent login request by a legitimate user.

\item \textbf{\textit{Session Key Disclosure}:} If $\mathcal{A}$ can compute the session key to establish a connection with the server, then a session key disclosure attack is applicable in the scheme. In the proposed mechanism, $U_{i}$ and $S_{j}$ individually compute the session key, $SK_{ij}=h(UID_{i}||ID_{j}||C_{i}||Loc_{j}||VT_{ij})$ using mutually agreed values. Thus, $\mathcal{A}$ requires $C_{i}$, $UID_{i}$, $Loc_{j}$, $VT_{ij}$, and $ID_{j}$ to compute the common session key. We consider that $\mathcal{A}$ gets $Loc_{j}$ and $ID_{j}$ as an internal adversary, but s/he cannot compute $C_{i}$, $UID_{i}$, and $VT_{ij}$ because s/he does not have essential credentials to calculate these values. Further, the session key is valid for $VT_{ij}$ in the proposed protocol. Thus, an adversary fails to calculate the session key exactly to perform malicious activities in the proposed scheme.

\item \textbf{\textit{Stolen Smart Card}:} In this attack, it is considered that $\mathcal{A}$ gets saved parameters in the user smart card ($SC_{i}$). If $\mathcal{A}$ can establish a connection with the server by sending a login request and computing a session key correctly to get services on behalf of a legitimate user, then a stolen smart card attack can be possible in the protocol. According to the proposed mechanism, $\mathcal{A}$ knows $W_{i}$, $X_{i}$, $Y_{i}$, $Z_{i}$, and $E_{i}$ by stealing $SC_{i}$ of $U_{i}$. Firstly, $\mathcal{A}$ should prove the legitimacy of $U_{i}$ as an adversary by computing $E^{\mathcal{A}}_{i}$, but s/he does not know $USID_{i}$, $USK_{i}$, and $PW_{i}$. Thus, $\mathcal{A}$ directly fails to justify the user authenticity in the first test. Moreover, $\mathcal{A}$ should compute $\alpha_{i}$ and $\beta_{i}$ to send a valid login request to $S_{j}$, but s/he cannot compute the necessary values as we have explained in a user impersonation attack (point-1 of Section 5). Besides, $\mathcal{A}$ is not able to calculate the common session key ($SK_{ij}$) due to unavailability of $UID_{i}$, $ID_{j}$, $C_{i}$, $Loc_{j}$. Hence, the proposed mechanism resists to a stolen smart card attack.

\item \textbf{\textit{Modification}:} If $\mathcal{A}$ can do any changes/updates in public channel values during the authentication and key agreement phase, then a modification attack is feasible in the system. To apply a modification attack in the proposed scheme, $\mathcal{A}$ should know $UID_{i}$, $ID_{j}$, $SSK_{j}$, $Loc_{j}$, $C_{i}$ because s/he needs to calculate $\alpha_{i}$, $\beta_{i}$, $\gamma_{i}$, $\sigma_{i}$ for the illegal update. However, $\mathcal{A}$ does not know all these required parameters, and s/he cannot also compute any values as a registered user or non-registered user. Thus, it makes difficult to do change(s) in any value of the proposed scheme.

\item \textbf{\textit{Password  Guessing}:} If $\mathcal{A}$ can check the correctness of a guessed password ($PW^{G}_{i}$) by comparing with the appropriate value, then a password guessing attack can be possible in the mechanism. To know the correctness of $PW^{G}_{i}$, $\mathcal{A}$ should compare the computed parameter with the appropriate value in which the original user password ($PW_{i}$) is used. In the proposed protocol, $PW_{i}$ is directly used in $A_{i}$, $B_{i}$, $X_{i}$, $Z_{i}$, $C_{i}$, and $E_{i}$ for the computation. $X_{i}$, $Z_{i}$, and $E_{i}$ are stored in the smart card ($SC_{i}$) of $U_{i}$. Thus, $\mathcal{A}$ should steal $SC_{i}$ anyhow to do the comparison, but we have discussed in point-3 (of Section 5) that a stolen smart card attack is not feasible in the proposed mechanism. Further, $A_{i}$ and $C_{i}$ are only known to $U_{i}$ and the registered servers. Further, it is not possible to compute $A_{i}$ and $D_{i}$ due to unavailability of $ID_{i}$, $USK_{i}$, $UID_{i}$, $K_{RC}$, $r_{3}$. Therefore, $\mathcal{A}$ has no opportunity to compare $PW^{G}_{i}$ with any computed parameter using $PW_{i}$. Hence, the proposed scheme is secure to a password guessing attack.

\item \textbf{\textit{Man-in-the-middle}:} When data is sent over a common communication channel and if an adversary can understand exchanged messages, then a man-in-the-middle attack is applicable in the system. $U_{i}$ sends $\alpha_{i}$, $\beta_{i}$, $\gamma_{i}$, $\sigma_{i}$ to $S_{j}$ over a public channel in the proposed authentication and key agreement phase. Therefore, $\mathcal{A}$ can capture these values to understand sent messages between $U_{i}$ and $S_{j}$. According to the proposed protocol, $\beta_{i}$, $\alpha_{i}$, and $\sigma_{i}$ are calculated as $h(UID_{i}||SSK_{j}||C_{i}||T_{1})$, $UID_{i} \oplus h(ID_{j}||SSK_{j}||T_{1})$, and $h(VT_{ij}||C_{i}||(T_{2}-T_{1}))$ respectively using one-way hash function. Hence, $\mathcal{A}$ cannot find any vital information or value, which can be used to understand the communication between $U_{i}$ and $S_{j}$, and these parameters are freshly computed for each communication session. $\gamma_{i}$ is calculated as $(VT_{ij}||Loc_{j}) \oplus h(C_{i}||UID_{i}||ID_{j}||\beta'_{i})$ and therefore, if $\mathcal{A}$ wants to reveal $(VT_{ij}||Loc_{j})$, then s/he should know $UID_{i}$, $C_{i}$, $ID_{j}$, $\beta'_{i}$, but $\mathcal{A}$ does not have all these essential parameters anyhow. Thus, the proposed scheme can withstand against a man-in-the-middle attack.

\item \textbf{\textit{Forward Secrecy}:} It is an attribute of key agreement mechanisms to guarantee that the session keys cannot be computed in the future even though session key long-term credentials are revealed. In the proposed scheme, the session key is calculated as $SK_{ij}=h(UID_{i}||ID_{j}||C_{i}||Loc_{j}||VT_{ij})$. Since an internal adversary knows $ID_{j}$ and $Loc_{j}$, s/he needs other values ($UID_{i}$, $C_{i}$, and $VT_{ij}$) for the session key computation. $UID_{i}$ is an anonymous identity of $U_{i}$, and it is computed as $h(r_{1}||ID_{i}||r_{2})$. $C_{i}$ is a computed parameter as $h(K_{RC}||r_{3}||A_{i}) \oplus USK_{i} \oplus h(UID_{i}||A_{i})$. Here, $C_{i}$ and $UID_{i}$ are long-term variables of $U_{i}$, but it is not feasible to get/compute these values anyhow due to unavailability of necessary values. Further, the required credentials (to compute $UID_{i}$ and $C_{i}$) are only known $U_{i}$, and they are not saved anywhere in the memory. $VT_{ij}$ is the validity of $SK_{ij}$, which is changed in every session key. Considering all the points, the proposed protocol satisfies forward secrecy.

\item \textbf{\textit{Replay}:} The prime motive to apply this attack is to delay/retransmit transmitted messages via a public channel later so users cannot get services on-time from the server. If an adversary succeeds to do these activities and the server/user accepts delayed messages in the system, then a replay attack is feasible. The time-stamp concept is used in the proposed scheme to compute $\alpha_{i}$, $\beta_{i}$, $\gamma_{i}$, and $\sigma_{i}$. When $S_{j}$ gets $\big\{\alpha_{i}, \beta_{i}, T_{1}\big\}$ from $U_{i}$, $S_{j}$ firstly checks its freshness through $\Delta T \leq T_{2} - T_{1}$, where $\Delta T$ is the threshold time, $T_{2}$ is the receiving time-stamp at $S_{j}$, and $T_{1}$ is the current time-stamp at $U_{i}$. If it holds, then only $S_{j}$ proceeds to the next step, else it is discarded directly. Similarly, $U_{i}$ confirms the validity of \big\{\text{$\gamma_{i}, \sigma_{i}, T_{2}$}\big\} at the receiving time-stamp ($T_{3}$). If $\mathcal{A}$ attempts to delay or resend previous messages, then it is immediately identified at the server/user side. Besides, all values are freshly computed for a new session, and therefore, it becomes difficult for $\mathcal{A}$ to deal with it. Thus, a replay attack is not possible in the proposed mechanism.

\begin{table*}[t]
\centering
\caption{Security Attributes Comparison among Various Authentication Schemes}
  \label{SComparison}		
  \begin{tabular}{lcccccccccccccc}
    \hline

		\hspace{0.25in} \textbf{Schemes} & \textbf{A1} & \textbf{A2} & \textbf{A3} & \textbf{A4} & \textbf{A5} & \textbf{A6} & \textbf{A7} & \textbf{A8} & \textbf{A9} & \textbf{A10} & \textbf{A11} & \textbf{A12} & \textbf{A13} & \textbf{A14} \\		\hline
		
		Odelu et al. \cite{Odelu2015} & $\boxtimes$ & $\checkmark$ & $\checkmark$ & $\checkmark$ & $\checkmark$ & $\checkmark$ & $\checkmark$ & $\checkmark$ & $\checkmark$ & $\checkmark$ & $\checkmark$ & $\checkmark$ & Yes & Yes \\ 

		Irshad et al. \cite{Irshad2017} & $\boxtimes$ & $\checkmark$ & $\checkmark$ & $\checkmark$ & $\checkmark$ & $\boxtimes$ & $\checkmark$ & $\checkmark$ & $\checkmark$ & $\checkmark$ & $\checkmark$ & $\checkmark$ & Yes & Yes \\ 
		
		Mishra et al. \cite{Mishra2018} & $\checkmark$ & $\boxtimes$ & $\checkmark$ & $\checkmark$ & $\checkmark$ & $\checkmark$ & $\checkmark$ & $\boxtimes$ & $\checkmark$ & $\checkmark$ & $\boxtimes$ & $\checkmark$ & Yes & Yes \\ 
		
		Qiao et al. \cite{Qiao2019} & $\checkmark$ & $\boxtimes$ & $\checkmark$ & $\boxtimes$ & $\checkmark$ & $\boxtimes$ & $\checkmark$ & $\boxtimes$ & $\boxtimes$ & $\checkmark$ & $\checkmark$ & $\checkmark$ & Yes & No \\ 
		
		Proposed & $\checkmark$ & $\checkmark$ & $\checkmark$ & $\checkmark$ & $\checkmark$ & $\checkmark$ & $\checkmark$ & $\checkmark$ & $\checkmark$ & $\checkmark$ & $\checkmark$ & $\checkmark$ & Yes & Yes \\ \hline
			
\end{tabular}
\end{table*}

\item \textbf{\textit{Insider}:} If any registered user ($U_{i}$) can get services from the server ($S_{j}$) on behalf of another legitimate user ($U_{i+1}$), then an insider attack can be launched in the system. In this case, $U_{i}$ is acting as an adversary for $U_{i+1}$ and a legitimate user. To get services on behalf of $U_{i+1}$, $U_{i}$ should compute $\alpha_{i+1}$ and $\beta_{i+1}$ correctly so $S_{j}$ considers that $\big\{\alpha_{i+1}, \beta_{i+1}, T_{1}\big\}$ is received from $U_{i+1}$ even though it is sent by $U_{i}$ for the illegal access of services. According to the proposed scheme, $\alpha_{i+1}$ and $\beta_{i+1}$ are calculated as $UID_{i+1} \oplus h(ID_{j}||SSK_{j}||T_{1})$ and $h(UID_{i+1}||SSK_{j}||C_{i+1}||T_{1})$ respectively. Thus, $U_{i}$ needs $UID_{i+1}$, $ID_{j}$, $SSK_{j}$, and $C_{i+1}$. $U_{i}$ is a registered user of the system and s/he knows his/her smart card parameters ($W_{i}$, $X_{i}$, $Y_{i}$, $Z_{i}$, $E_{i}$). Thus, $U_{i}$ can reveal $ID_{j}$ and $SSK_{j}$ from $List_{S_{j}}$ ($=h(ID_{i}||PW_{i}||r_{2}) \oplus h(r_{1}||ID_{i}||PW_{i}) \oplus Z_{i}$), but s/he is not able to get $UID_{i+1}$ and $C_{i+1}$ because they are calculated as $h(r_{1}||ID_{i+1}||r_{2})$ and $X_{i+1} \oplus h(r_{2} \oplus ID_{i+1}) \oplus h(r_{1} \oplus PW_{i+1})$ respectively. Here, $r_{1}$ and $r_{2}$ are random nonce, and they are not the same as selected by $U_{i}$ during the registration phase. Further, $ID_{i+1}$ and $PW_{i+1}$ are the identity and password of $U_{i+1}$, and they are not known to anyone. Thus, $U_{i}$ cannot compute all required parameters to forge $U_{i+1}$ even though s/he is a registered user in the system. Hence, the proposed scheme is secure to an insider attack.

\item \textbf{\textit{Denial of Service}:} An adversary aims to reduce the system performance by sending multiple requests so that the receiver becomes busy to verify the received requests. Thus, s/he can launch a DoS attack to achieve his/her intentions. The proposed system is focused on multi-server architecture to offer services from different authorized servers. Therefore, users can connect with different servers for quick services. Besides, a server ($S_{j}$) confirms the validity of the obtained requests through $\Delta T$. Thus, if any request is delayed by $\mathcal{A}$, then it is identified immediately. Furthermore, if $\mathcal{A}$ sends multiple requests with different time-stamps to clear the $\Delta T$ test, then also $S_{j}$ can distinguish an illegal action quickly through $\beta'_{i}\stackrel{?}{=}\beta_{i}$ because the original time-stamp ($T_{1}$) is used in $\beta_{i}$ computation, and $\beta'_{i}$ is computed based on the received request time-stamp (which is different because $\mathcal{A}$ has changed it.). Moreover, $S_{j}$ requires to execute only two SHA-256 operations, and the execution time of SHA-256 is comparatively less. Hence, the proposed protocol promptly identities a DoS attack and resists against it.

\end{enumerate}

Table \ref{SComparison} shows the comparison of different security attributes for various relevant multi-server architecture based authentication mechanisms. We have denoted different security attacks and attributes in Table \ref{SComparison} as \textbf{A1:} User impersonation\textbf{;} \textbf{A2:} Server impersonation\textbf{;} \textbf{A3:} Session key disclosure\textbf{;} \textbf{A4:} Stolen smart card\textbf{;} \textbf{A5:} Modification\textbf{;} \textbf{A6:} Password guessing\textbf{;} \textbf{A7:} Man-in-the-middle\textbf{;} \textbf{A8:} Replay\textbf{;} \textbf{A9:} Insider\textbf{;} \textbf{A10:} Forward Secrecy\textbf{;} \textbf{A11:} Denial of service\textbf{;} \textbf{A12:} user anonymity \textbf{;} \textbf{A13:} Mutual authentication\textbf{;} \textbf{A14:} user password update\textbf{;}~~$\checkmark:$ Resists\textbf{;}~~$\boxtimes:$ Vulnerable\textbf{;}

\section{Performance Results of the Proposed Scheme}
We discuss performance results for various measures such as computational time, storage cost, and communication cost. Further, the performance results are compared with relevant multi-server authentication schemes.

\subsection{Execution Cost}
The execution cost is defined as the total number of required different cryptographic operations during the authentication and key agreement scheme. These operations require some amount of time for the implementation, which is called as the execution time, and it is measured is milliseconds (\textit{ms}) \cite{Limbasiya2020}. Relevant authentication schemes are mainly designed using different cryptographic functions namely one-way hash ($T_{h(\cdot)}$), elliptic curve multiplication ($T_{ECM}$), Chebyshev chaotic ($T_{CC}$), Symmetric key cryptography ($T_{AES}$), bit-wise XOR ($T_{\oplus}$) and concatenation ($T_{||}$). We consider the system configuration as Ubuntu 18.04 64-bit operating system, 8 GB RAM, and Intel 2.4 GHz CPU to measure the average execution time of used cryptographic operations. We have implemented all these operations on the configured platform using Python libraries (Pycrypto, fastecdsa, and hashlib) to get the average execution time after 100 runs. The execution time of $T_{\oplus}$ and $T_{||}$ is minimal compared to $T_{h(\cdot)}$, $T_{CC}$, $T_{ECM}$, and $T_{AES}$. Table \ref{exec} shows the individual execution time for each operation. Thus, we consider the execution time of $T_{ECM}$, $T_{h(\cdot)}$, $T_{AES}$, and $T_{CC}$ to calculate the execution time of the proposed scheme, \cite{Odelu2015}, \cite{Irshad2017}, \cite{Mishra2018}, and \cite{Qiao2019}.

\begin{table}[!h]
\centering
\caption{The Average Execution Cost of Various Operations}
  \label{exec}		
  \begin{tabular}{cc}
    \hline
		\textbf{Operation} & \textbf{Execution Time}  \\ \hline
		$T_{h(\cdot)}$ & 0.0006 \textit{ms}\\
		$T_{CC}$ & 21.0400 \textit{ms} \\
		$T_{ECM}$ & 1.3870 \textit{ms} \\
		$T_{AES}$ & 0.0012 \textit{ms} \\
\hline		
\end{tabular}
\end{table}

\begin{table*}[h]
\centering
\caption{Execution Cost and Time Statistics for Relevant Multi-server Authentication Schemes}
  \label{Execution}		
  \begin{tabular}{lcc}
    \hline
	& \multicolumn{2}{c}{\textbf{Execution Cost}}\\ \cline{2-3}
		\hspace{0.25in} \textbf{Schemes} & \textbf{Registration} & \textbf{Authentication and Key Agreement} \\		\hline
		
		Odelu et al. \cite{Odelu2015} & 6$T_{h(\cdot)}$~+~1$T_{ECM}$ ($\approx$ 1.3906 \textit{ms}) & 16$T_{h(\cdot)}$~+~6$T_{AES}$~+~2$T_{ECM}$ ($\approx$ 2.7908 \textit{ms}) \\ 

		Irshad et al. \cite{Irshad2017} & 3$T_{h(\cdot)}$ ($\approx$ 0.0018 \textit{ms}) & 29$T_{h(\cdot)}$~+~6$T_{CC}$ ($\approx$ 126.2574 \textit{ms}) \\ 
		
		Mishra et al. \cite{Mishra2018} & 6$T_{h(\cdot)}$ ($\approx$ 0.0036 \textit{ms}) & 19$T_{h(\cdot)}$~+~7$T_{ECM}$ ($\approx$ 9.7204 \textit{ms}) \\ 
		
		Qiao et al. \cite{Qiao2019} & 2$T_{h(\cdot)}$~+~1$T_{CC}$~+~1$T_{AES}$ ($\approx$ 21.0424 \textit{ms}) & 6$T_{h(\cdot)}$~+~4$T_{CC}$~+~2$T_{AES}$ ($\approx$ 84.166 \textit{ms})\\ 
		
		Proposed & 20$T_{h(\cdot)}$ ($\approx$ 0.012 \textit{ms}) & 20$T_{h(\cdot)}$ ($\approx$ 0.012 \textit{ms})\\
\hline		
\end{tabular}
\end{table*}

In the authentication scheme, the registration phase is executed once only, but the authentication and key agreement phase is performed whenever a user wants to get services from the server. Hence, it is indispensable to have less execution cost in the authentication and key agreement phase than the registration phase. Table \ref{Execution} describes the execution cost/time for recent schemes \cite{Odelu2015}, \cite{Irshad2017}, \cite{Mishra2018}, \cite{Qiao2019}, and the proposed protocol. We notice that the execution cost is very high in \cite{Irshad2017} and \cite{Qiao2019} during the authentication and key agreement phase due to the usage of high-cost operation ($T_{CC}$). Moreover, the execution time is comparatively high in \cite{Odelu2015} and \cite{Mishra2018} because they specifically used ECC in the protocol design, and its execution time is high compared to the execution time of SHA-256. However, the proposed scheme takes very less time for the implementation compared to \cite{Odelu2015}, \cite{Irshad2017}, \cite{Mishra2018}, and \cite{Qiao2019}.

\subsection{Communication and Storage Costs}
When both (user and server) want to establish a connection with each other for some services, they exchange different parameters for mutual authentication during the authentication and key agreement phase. The cost of these different parameters is called the communication cost, and it is measured in bytes due to memory requirements. During the registration phase, some computed values are saved in the user smart card, which are used in future computations to verify the user and compute the login request. The cost of saved values is called the storage cost, and it is measured in bytes \cite{Limbasiya2020}. In general, an identity/normal variable/time-stamp needs 8 bytes, Chebyshev chaotic requires 16 bytes, SHA-256 needs 32 bytes, elliptic curve (EC) needs 64 bytes, a time-stamp needs 4 bytes, and AES symmetric encryption requires 32 bytes for communication \cite{Qiao2019}, \cite{Limbasiya2020}. 

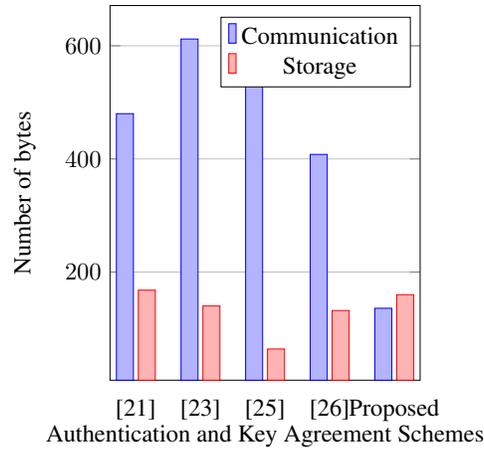
\begin{figure}[!h]
\centering \scalebox{0.92}{
\begin{tikzpicture}
\begin{axis}
[ width  = 0.5\textwidth, height = 7cm, major x tick style =
transparent, ybar, bar width=7pt, ymajorgrids = true,
xlabel={Authentication and Key Agreement Schemes},
ylabel = {Number of bytes}, symbolic x coords={\cite{Odelu2015}, \cite{Irshad2017}, \cite{Mishra2018}, \cite{Qiao2019}, Proposed}, xtick = data, scaled y ticks = false, legend pos=north east]

\addplot coordinates {(\cite{Odelu2015},480) (\cite{Irshad2017},612) (\cite{Mishra2018},616) (\cite{Qiao2019},408) (Proposed,136)};   

\addplot coordinates {(\cite{Odelu2015},168) (\cite{Irshad2017},140) (\cite{Mishra2018},64) (\cite{Qiao2019},132) (Proposed,160)}; 
 
\legend{Communication, Storage}  

\end{axis}
\end{tikzpicture}}
\caption{Communication and Storage Cost Comparison for Relevant Authentication Mechanisms}
\label{COST}
\end{figure}

In \cite{Odelu2015}, it needs 4 (AES), 11 (one-way hash) during the communication and 5 (one-way hash), 1 (identity) for the storage. The scheme \cite{Irshad2017} requires 16 (one-way hash), 4 (Chebyshev chaotic), 1 (identity), 3 (time-stamp) for communication and 3 (one-way hash), 1 (Chebyshev chaotic), 3 (identity) for storage. Mishra et al. \cite{Mishra2018} needs 11 (one-way hash), 4 (ECC), 2 (time-stamp) in communication and 2 (one-way hash) as the storage cost. The protocol \cite{Qiao2019} requires 5 (one-way hash), 2 (ECC), 6 (Chebyshev chaotic) during communication and 2 (one-way hash), 1 (AES), 1 (Chebyshev chaotic), 2 (identity) for the storage. The proposed scheme needs 4 (one-way hash), 2 (time-stamp) as the communication cost, and 5 (one-way hash) as the storage cost. We have calculated the communication and storage costs for each authentication protocol, and their comparison is shown in Fig. \ref{COST}. It is observed that the storage cost is more in the proposed scheme compared to \cite{Irshad2017}, \cite{Mishra2018}, and \cite{Qiao2019}, but it is a one-time cost. The communication cost is required whenever both (user and server) are interested in establishing a connection. Thus, the communication cost should be less in the system, and we can observe in Fig. \ref{COST} that the communication cost is very less compared to \cite{Odelu2015}, \cite{Irshad2017}, \cite{Mishra2018}, and \cite{Qiao2019}.

\section{Conclusion}
We have proposed a privacy-preserving authentication and key agreement protocol for a multi-server healthcare system using only SHA-256, bit-wise XOR, and concatenation operations. The proposed scheme resists to stolen smart card, insider, password guessing, man-in-the-middle, user/server impersonation, forward secrecy, session key disclosure, denial of service, and modification attacks, achieving various privacy and security properties for medical users. Moreover, the performance results are also efficient in terms of the execution time, communication overhead, and storage cost. Hence, the proposed protocol achieves security and privacy requirements while taking comparatively less computational resources than relevant mechanisms. Therefore, the proposed scheme helps multi-server healthcare systems, protecting user data and privacy with less computational resources.

\end{document}